\documentclass[preprint,12pt,longnamesfirst]{aastex}

\shorttitle{Optical and X-ray Study of SNR Candidates in 30 Dor}
\shortauthors{Chu et al.}

\begin{document}


\title{An Optical and X-ray Examination of Two Radio Supernova 
Remnant Candidates in 30 Doradus}


\author{
You-Hua Chu\altaffilmark{1,2}, Robert A.\ Gruendl\altaffilmark{1}, 
C.-H.\ Rosie Chen\altaffilmark{1},\\ Jasmina S. Lazendic\altaffilmark{3},
and John R.\ Dickel\altaffilmark{1}}


\altaffiltext{1}{Department of Astronomy, University of Illinois at 
Urbana-Champaign, 1002 West Green Street, Urbana, IL 61801}
\altaffiltext{2}{
Visiting Astronomer, Cerro Tololo Inter-American Observatory,
National Optical Astronomy Observatories, operated by the
Association of Universities for Research in Astronomy, Inc.\ (AURA)
under a cooperative agreement with the National Science Foundation.}
\altaffiltext{3}{
Harvard-Smithsonian Center for Astrophysics, 60 Garden Street, 
Cambridge, MA 02138}
\email{chu@astro.uiuc.edu, gruendl@astro.uiuc.edu,c-chen@astro.uiuc.edu,
jlazendic@cfa.harvard.edu, johnd@astro.uiuc.edu}

\begin{abstract}

The giant \ion{H}{2} region 30 Doradus is known for its violent
internal motions and bright diffuse X-ray emission, suggesting 
the existence of supernova remnants (SNRs), but no nonthermal
radio emission has been detected.
Recently, Lazendic et al.\ compared the H$\alpha$/H$\beta$ 
and radio/H$\alpha$ ratios and suggested two small radio sources
to be nonthermal and thus SNR candidates; however, no optical
or X-ray counterparts were detected.
We have used high-resolution optical images and high-dispersion 
spectra to examine the morphological, spectral, and kinematic
properties of these two SNR candidates, and still find no 
optical evidence supporting their identification as SNRs. 
We have also determined the X-ray luminosities of these SNR 
candidates, and find them 1--3 orders of magnitude lower than 
those commonly seen in young SNRs.
High extinction can obscure optical and X-ray signatures of
an SNR, but would prohibit the use of a high radio/H$\alpha$ 
ratio to identify nonthermal radio emission.
We suggest that the SNR candidate MCRX J053831.8$-$690620
is associated with a young star forming region; while
the radio emission originates from the obscured star 
forming region, the observed optical emission is dominated
by the foreground.
We suggest that the SNR candidate MCRX J053838.8$-$690730 
is associated with a dust/molecular cloud, which obscures 
some optical emission but not the radio emission.

\end{abstract} 


\keywords{
\ion{H}{2} regions --- ISM: individual (30 Doradus) --- Magellanic 
Clouds --- supernova remnants --- dust, extinction}

\newpage

\section{Introduction}

The giant \ion{H}{2} region 30 Doradus is a site of active star 
formation.
A range of stellar ages are present in 30 Dor \citep{WB97}, from
enshrouded proto-stars \citep{Retal98,Wetal99,Wetal02,Betal01} 
through the 2-5 Myr old R136 cluster \citep{Hetal95, Setal99} to 
the 20-25 Myr old Hodge 301 cluster \citep{GC00}. 
The powerful ionizing flux from the young prominent R136 cluster 
makes 30 Dor the most luminous \ion{H}{2} region in the entire Local
Group.  
Fast stellar winds from the current-generation massive stars and 
supernova explosions from the former-generation massive stars have 
accelerated the interstellar medium (ISM) to high velocities 
with $\Delta v$ = 100--300 km~s$^{-1}$ and heated the ISM to 
X-ray-emitting temperatures \citep{CK94, Wang99}.
While the observed diffuse X-ray emission and violent gas motion 
provide ample evidence for the existence of supernova remnants (SNRs)
in 30 Dor, radio observations were unable to confirm the SNRs through
spectral indices because of the overwhelming background thermal 
emission.

The existence of nonthermal radio emission superposed on thermal
emission may be diagnosed from comparisons between radio continuum
and hydrogen Balmer line emission.  
In \ion{H}{2} regions the radio free-free emission and optical
recombination line emission orginate from different mechanisms
but make use of a common population of free electrons,
while in SNRs the radio synchrotron emission and the optical
recombination lines originate from both different materials
and different mechanisms.
The ratio of radio continuum to recombination line emission 
is much higher in SNRs than in \ion{H}{2} regions; therefore, 
the existence of nonthermal radio emission can be diagnosed
by a high radio/H$\alpha$ flux ratio if the foreground 
extinction is low, e.g., N66 in the Small Magellanic Cloud 
\citep{Yetal91}.

Using a 13-cm radio map and an H$\alpha$ image of 30 Dor, 
\citet{Detal94} found two regions with radio/H$\alpha$ 
ratios 20 times higher than average and suggested that they 
represented an extra extinction of 3.3 mag.
Recently, \citet{LDJ03} re-examined the radio/H$\alpha$ 
ratios with a higher angular resolution at 3 cm and 6 cm,
and used H$\alpha$ and H$\beta$ images to make an independent
assessment of the extinction.
They identified four regions with high radio/H$\alpha$ ratios 
but low H$\alpha$/H$\beta$ ratios or weak H$\beta$ emission.
Two of these regions, corresponding to the ones identified by 
\citet{Detal94}, were suggested to be SNR candidates;
the remaining two were suggested to be young \ion{H}{2} regions
because they were coincident with previously identified
proto-stellar objects \citep{LDJ03}.
Note however that 30 Dor consists of a complex mixture of gas
and dust and that the observed optical emission is inevitably 
biased toward regions on the near side of dust clouds; 
therefore, the extinction determined from the H$\alpha$/H$\beta$
ratios may not represent that of radio sources behind the dust.
It is thus necessary to ascertain the co-spatiality of the radio
and optical emission for SNR candidates identified by high 
radio/H$\alpha$ and low H$\alpha$/H$\beta$ ratios, and to critically
assess their nature at other wavelengths.

We will refer to the two radio SNR candidates in 30 Dor, 
MCRX J053831.8$-$690620 and MCRX J053838.8$-$690730, as
candidates 1 and 2, respectively, throughout the rest of the paper.
These two SNR candidates are small; their radio sizes, 
10$''$ and 8$''$, correspond to 2.5 and 2 pc at the Large 
Magellanic Cloud (LMC) distance of 50 kpc.
They show compact rather than shell structures at radio 
wavelengths and show no bright X-ray emission over the background.
These properties are uncommon among known SNRs in the LMC or the
Galaxy, but are similar to the compact radio SNR candidates found 
in distant starburst galaxies, such as M82 \citep{Metal94}.
\citet{LDJ03} argued that the complex interstellar environment in
a giant \ion{H}{2} region smothered the other SNR 
signatures\footnote{\citet{LDJ03} misprinted the UV absorption lines
reported by \citet{Cetal94} as forbidden lines in their \S 4.1.}.

We have analyzed high-resolution optical images and spectra
of the two radio SNR candidates in 30 Dor, but found no supporting 
evidence for their identification as SNRs.
We have also derived their X-ray luminosities and find values low 
for young shell-type SNRs.
This paper reports our optical and X-ray examination of the two 
radio SNR candidates in 30 Dor: \S2 describes the observations
used in our analysis, \S3 discusses the optical morphologies, 
spectral properties, and kinematics of these SNR candidates; 
\S4 provides an X-ray test of the SNR candidacy, and \S5 
summarizes our conclusions.

\section{Optical and X-ray Observations}

We have made use of archival {\it Hubble Space Telescope (HST)} WFPC2
images of 30 Dor taken with the H$\alpha$ $\lambda$6563 
(F656N), [\ion{O}{3}] $\lambda$5007 (F502N), and [\ion{S}{2}] 
$\lambda\lambda$6717, 6731 (F673N) filters on 1994 January 2
with a total exposure time of 1000 s in each filter.
These images were previously reported in detail by 
\citet{Setal98}.
We processed and calibrated these data using standard 
IRAF/STSDAS procedures as described by \citet{Cetal00}.
In addition, we have used the filter curves to determine
the transmission appropriate for the $\sim$300 km~s$^{-1}$
red-shift of 30 Dor and used spectrophotometrically 
determined [\ion{N}{2}]/H$\alpha$ ratios \citep{MCP85}
to determine the contamination of the [\ion{N}{2}] $\lambda$6548
line to the H$\alpha$ image.  
No corrections are needed for the [\ion{O}{3}] and [\ion{S}{2}]
fluxes because the F502N and F673N filters are flat, but a
net correction of +6\% is needed for the H$\alpha$ flux owing
to the sloping filter curve and the [\ion{N}{2}] contamination.

We have also used high-dispersion long-slit spectra obtained 
with the echelle spectrograph on the 4m telescope at Cerro 
Tololo Inter-American Observatory to study kinematic properties
of the SNR candidates.
The echelle observations of SNR candidate 1 were
taken on 1995 January 20 at offsets of 15$''$ and 18$''$ 
south of R136; the instrumental setup has been described
by \citet{Petal99}.  Roughly, the data cover both the 
H$\alpha$ and [\ion{N}{2}] $\lambda\lambda$6548, 6584
lines with an instrumental FWHM of 16 km~s$^{-1}$ and 
an image scale of 3.75 km~s$^{-1}$ pixel$^{-1}$ along the 
dispersion and 0\farcs267 pixel$^{-1}$ along the slit.
The observation that covered SNR candidate 2 was obtained
on 1988 January 8 and belonged to the data set used 
by \citet{CK94} to study the kinematic structure of 30 Dor.
This earlier observation used a different camera 
and CCD, resulting in an instrumental FWHM of 18 km s$^{-1}$
and an image scale of 9.6 km~s$^{-1}$ pixel$^{-1}$ along
the dispersion and 0\farcs635 pixel$^{-1}$ along the slit.
A slit width of 250 $\mu$m (1\farcs64) was used for all
echelle observations.

An archival {\it Chandra} ACIS-I GTO observation 
(Obs ID: 62520; PI: G. Garmire) of 30 Dor was also used in 
this paper.  
The observation was made on 1999 September 21 for an
exposure time of 25.5 ks.
The data were kindly provided to us by L.\ Townsley.
Details of this observation will be reported in a
paper by Townsley et al.\ (2004, in preparation).

\section{Morphological, Spectral, and Kinematic Properties}

Optical signatures of SNRs vary according to their origins, 
interstellar and stellar environments, and evolutionary stages.
We will discuss the physical significance of the morphological
and spectral signatures of SNRs and use the archival {\it HST} 
WFPC2 images to assess the nature of SNR candidates 1 and 2 
in 30 Dor.  
We will also use the echelle spectra to search for high-velocity 
features as an indication of SNR shocks.

Figure 1 shows {\it HST} WFPC2 images of the two SNR candidates
in the H$\alpha$, [\ion{S}{2}] $\lambda\lambda$6716, 6731, and
[\ion{O}{3}] $\lambda$5007 lines.
No distinct nebular features, such as shells, are seen at the 
positions of the SNR candidates.
An ionized SNR shell morphology is visible only if the ambient
medium is sufficiently dense in most directions.
The threshold ambient hydrogen density $n_0$ can be estimated 
from the detection limit of the emission measure, 
$<n_{\rm e}^2> \ell$, where $n_{\rm e}$ is the electron density 
of the SNR shell material and $\ell$ is 
the line-of-sight length of the emitting region.
Assuming an adiabatic SNR shock (so that $n_{\rm e} \sim 4 n_0$)
and adopting the SNR radius as an order-of-magnitude 
approximation for $\ell$, the threshold ambient density becomes 
$n_0 \sim$ 0.25 (emission measure/radius)$^{1/2}$.
The WFPC2 H$\alpha$ image has been used to measure surface
brightnesses.
In the region of SNR candidate 1, the H$\alpha$ surface brightness
is $5.9\times10^{-14}$ ergs~cm$^{-2}$~s$^{-1}$~arcsec$^{-2}$, and
the 3$\sigma$ detection limit above this background is 
$2.3\times10^{-14}$ ergs~cm$^{-2}$~s$^{-1}$~arcsec$^{-2}$.
In the region of SNR candidate 2, the H$\alpha$ surface brightness
is $1.4\times10^{-14}$ ergs~cm$^{-2}$~s$^{-1}$~arcsec$^{-2}$, and 
the 3$\sigma$ detection limit above this background is
$1.4\times10^{-14}$ ergs~cm$^{-2}$~s$^{-1}$~arcsec$^{-2}$.
As the H$\alpha$ surface brightness SB(H$\alpha$) = 
$1.9\times10^{-18}~n_{\rm e}^2~\ell$ 
ergs~cm$^{-2}$~s$^{-1}$~arcsec$^{-2}$ for $n_{\rm e}$ in units
of cm$^{-3}$ and $\ell$ in pc, 
these 3$\sigma$ detection limits correspond to emission measures
of $1.2\times10^4$ and $7.4\times10^3$ cm$^{-6}$ pc, respectively.
The SNR candidates have radii of 1.25 pc and 1.0 pc, and thus
both have a threshold $n_0$ of $\sim$ 25 H-atom cm$^{-3}$.
If these candidates were SNRs in a dense medium with
$n_0 \ge 10^2$ H-atom cm$^{-3}$ as derived by \citet{LDJ03} 
from their sizes, they should have shown visible shell 
structures in the {\it HST} WFPC2 H$\alpha$ images displayed 
in Figure 1.
The implication of the lack of H$\alpha$ shell morphology
depends on whether the optical emission is co-spatial with 
the radio sources or not.
If they are co-spatial, the lack of visible shell structure
raises doubts as to the SNR nature of the candidates.
On the other hand, if the optical emission is dominated by a 
foreground component, then the H$\alpha$/H$\beta$ ratio is
irrelevant to the interpretation of the radio sources.
In either case, the existence of SNRs is not justified.

An SNR can be spectroscopically diagnosed by (1) a high  
[\ion{S}{2}]/H$\alpha$ ratio, e.g., $\ge$0.45 \citep{MF97}, 
if the SNR is not in a hard ultraviolent radiation field
and the SNR is old enough for sulfur to become collisionally 
ionized to S$^{+}$ in the post-shock region; (2) bright 
broad oxygen lines with no hydrogen Balmer line counterparts,
if the SNR is young and dominated by O-rich supernova ejecta 
\citep[e.g.,][]{KB80}; or (3) hydrogen Balmer lines with 
an absence of forbidden lines such as [\ion{O}{1}], 
[\ion{O}{3}], and [\ion{S}{2}], if the SNR is produced by
a Type Ia supernova and the SNR interacts with a neutral
ISM \citep{CKR80}.
Using the {\it HST} WFPC2 images, we have measured 
the H$\alpha$, [\ion{O}{3}], and [\ion{S}{2}] fluxes
within 10$''$- and 8$''$-diameter apertures that 
correspond to the radio source sizes of SNR candidates
1 and 2.
We find [\ion{S}{2}]/H$\alpha$ ratios of 0.09$\pm$0.01 
and 0.10$\pm$0.01 and [\ion{O}{3}]/H$\alpha$ ratios of 
1.1$\pm$0.1 and 0.51$\pm$0.04 for SNR candidates 1 and 2,
respectively.
Similar line ratios are present in surrounding regions,
and these ratios are typical for an \ion{H}{2} region 
photoionized by early-type O stars.
Therefore, the optical emission toward the two SNR candidates
does not show line ratios indicative of SNRs.

\citet{CK86,CK88,CK94} have demonstrated that SNRs can be 
effectively diagnosed kinematically by detections of 
shocked material at large velocity offsets from the systemic
velocities, $\Delta v >$ 100 km~s$^{-1}$, even if the SNRs 
are embedded in giant \ion{H}{2} regions.
In Figure 2, we show three east-west oriented long-slit 
echelle spectrograms of the H$\alpha$ line; two intersect
SNR candidate 1 and one intersects SNR candidate 2.
It is evident in Figure 2 that high-velocity features are
present in the X-ray-bright cavity that encompasses the R136 
cluster, but not in the vicinity of the SNR candidates.
The nebular kinematics do not show shocks to support the
SNR nature of these two candidates.

\citet{LDJ03} cited a lower electron temperature and higher 
helium abundance reported by \citet{RM87} in region 4B near 
candidate 2 to support the existence of supernova ejecta;
however, this anomalous temperature and abundance was not
confirmed by \citet{LT91}.
These latter authors further noted that this region was
``particularly obscured".
Thus, we do not find any optical morphological, spectral, or
kinematic properties of candidates 1 and 2 to support their
identification as SNRs.

The suggestion of nonthermal radio emission from candidate 2 
was mainly based on the high radio/H$\alpha$ ratio,
as the H$\beta$ emission was very weak \citep{LDJ03}.
The {\it HST} WFPC2 images in Figure 1 and in the top panel 
of Figure 2 show that SNR candidate 2 is coincident with a
portion of an irregularly shaped dust cloud, which appears 
as a  void in the H$\alpha$ image \citep{LDJ03}.
The H$\alpha$ echelle spectrogram in the bottom panel of Figure 2 
shows the same two velocity components in the low surface 
brightness region of candidate 2 and in its neighboring regions, 
at heliocentric velocities of $\sim$280 and 220 km~s$^{-1}$.
The continuation of kinematic features suggests that the lower
surface brightness at candidate 2 is caused by extinction 
arising from a foreground dust cloud, which is clearly visible 
in the {\it HST} WFPC2 images.
The H$\alpha$ surface brightness in the dusty region is a 
factor of 10--20 lower than those of the neighboring regions 
for the 280 km~s$^{-1}$ component, and 5--9 times lower for 
the 220 km~s$^{-1}$ component.
If the intrinsic surface brightness behind the dusty
feature is similar to those of the neighboring regions,
then the extra extinction is $A_{\rm H \alpha}$ = 2.5--3.3
for the 280 km~s$^{-1}$ component and $A_{\rm H \alpha}$ = 
1.8--2.3 for the 220 km~s$^{-1}$ component.
As the 280 km~s$^{-1}$ component is the brighter of the two,
its extinction is a better approximation of the effective 
extinction of the total emission.
Indeed, the extinction of the 280 km~s$^{-1}$ component is
consistent with the 3.3 mag extra extinction implied by the
radio/H$\alpha$ ratio \citep{Detal94,LDJ03}.
Therefore, we conclude that SNR candidate 2 corresponds to a 
dusty region with an extra extinction of $\sim$ 3 mag at the 
H$\alpha$ line, and that there is no compelling evidence for
nonthermal radio emission.

\section{X-ray Test of the SNR Candidacy}

The two SNR candidates in 30 Dor do not have corresponding 
X-ray sources \citep{LDJ03,Wang99}. This is uncommon for young
SNRs found in the Galaxy or the LMC, which are usually very
luminous, particularly the ones with massive supernova
progenitors; their X-ray luminosities frequently reach
10$^{37}$ ergs~s$^{-1}$ \citep{Metal83}.

We re-examined the archival {\em Chandra} ACIS-I observation 
62520 centered at R136 \citep{townsley02,LDJ03}.
To extract X-ray count rates for the two SNR candidates,
we use a 5$''$-radius source aperture centered at the
radio position of each object and a 100$''$-radius background
aperture centered on an emission-free region at 
$\alpha$(J2000) = 05$^h$39$^m$56\fs41, 
$\delta$(J2000) = $-$69$^\circ$12$'$21\farcs3.
We find background-subtracted count rates of 
$6.0 \pm 2.5 \times10^{-4}$ and $9.5 \pm 3.0\times10^{-4}$ 
counts s$^{-1}$ for candidates 1 and 2, respectively.
To convert these count rates to X-ray luminosities, we need
to make assumptions of spectral shapes and foreground absorption.
For spectral shapes, we have used  \citet{RS77} models for
thin plasma emission with 0.4 solar abundances at temperatures 
of $kT$ = 0.5, 1.0, and 5.0 keV, as well as power-law models 
with a photon index of $\Gamma$ = 2.0 for pulsars and 
pulsar-wind nebulae \citep[e.g., SNR 0540$-$69.3,][]{Ketal01}.
For the foreground absorption, we assume visual extinctions of 
$A_V$ = 1, 3, 6, and 10 and adopt an LMC gas-to-dust ratio of
$N_{\rm H} = 7.5\times10^{21} A_V$ H-atom cm$^{-2}$ 
\citep{Fitz86}. 
The X-ray luminosity ($L_{\rm X}$) in the 0.5--10 keV band 
derived for these different models are listed in Table 1.

While young SNRs are known for their high X-ray luminosity, 
10$^{36}$--10$^{37}$ ergs~s$^{-1}$ \citep{Metal83,Wi99},
the X-ray luminosities of the two radio SNR candidates in 
30 Dor are 1--3 orders of magnitude lower. 
Their X-ray luminosities can be raised to 10$^{36}$ ergs~s$^{-1}$
if the foreground extinction is higher than $A_V = 10$, but such a 
high extinction would produce a high radio/H$\alpha$ ratio and
remove the basis for the initial identification of nonthermal 
radio emission and abrogate the SNR candidacy.

Compact radio SNR candidates without X-ray counterparts have 
been reported in starburst galaxies 
\citep[e.g., M82;][]{Metal94,Getal00}.
The combination of high radio and low X-ray luminosities
is expected for radiative SNRs expanding in the interclump medium
of molecular clouds.
As modeled by \citet{CF01}, the X-ray luminosity of an SNR in a 
dense medium peaks before the radiative phase and remains 
$\ge$ 10$^{37}$ ergs~s$^{-1}$ for up to 10$^3$ yr.
A radiative SNR in a dense medium is faint in X-rays, but should
be a strong emitter of optical forbidden lines.
High extinctions in starburst galaxies have prevented the
detection of optical counterparts of the compact radio SNR 
candidates.
Such high extinction might exist in 30 Dor but again would
invalidate the deduction of nonthermal radio emission from 
high radio/H$\alpha$ ratios.
Therefore, the absence of bright X-ray emission does not support
the SNR nature of the radio SNR candidates in 30 Dor.

\section{Summary and Conclusions}

Based on comparisons of H$\alpha$/H$\beta$ and radio/H$\alpha$
ratios \citet{LDJ03} suggested two radio SNR candidates in 30 Dor.
We have examined their morphological, spectral, and 
kinematic properties at optical wavelengths, but do not
find supporting evidence for their identification as SNRs.
We have further examined their X-ray properties, and find 
their low X-ray luminosities inconsistent with young SNRs.
Large extinction could obscure the optical and X-ray 
emission from a hypothetical SNR, but such large extinction 
would produce high H$\alpha$/H$\beta$ and radio/H$\alpha$
ratios which contradict the observations and weaken the 
argument for nonthermal radio emission.
Therefore, there is no compelling evidence that either
of the radio SNR candidates are truly SNRs.

An indication of the true nature of the two radio SNR
candidates in 30 Dor is provided by the recent {\it Spitzer 
Space Telescope (SST)} observations.
Bright IR objects in massive star forming regions have been 
associated with proto-stellar objects ever since the pioneering
observation of the BN object in Orion by \citet{BN67}. 
A press-release {\it SST} IRAC color composite image of 30 
Dor\footnote{Available at
http://www.spitzer.caltech.edu/Media/releases/ssc2004-01/ssc2004-01a.shtml.}
shows enhanced mid-IR emission coincident with both radio
SNR candidates as well as the two high radio/H$\alpha$ sources 
that were associated with proto-stellar objects by \citet{LDJ03}.
These two latter sources, MCRX\,J053848.3$-$690442 and 
MCRX\,J053848.6$-$690412, have been clearly demonstrated by 
{\it HST} NICMOS observations to contain compact embedded
young clusters \citep{Betal01,Wetal99,Wetal02}.

SNR candidate 1 is coincident with molecular clouds 30Dor-12
and 30Dor-13 \citep{Jetal98}, and furthermore is near two
Class I protostars, 30Dor-NIC07a and 30Dor-NIC07b
\citep{Betal01}.
Therefore, we suggest that candidate 1 is also associated
with a young star forming region; while the observed radio
emission originates from the embedded star forming region,
the observed optical emission arises from the foreground.

SNR candidate 2 shows diffuse mid-IR emission, and is 
coincident with a visible dust cloud (Figs.\ 1 \& 2)
and the molecular cloud 30Dor-15 \citep{Jetal98}.
Its radio/H$\alpha$ ratio suggests an extra extinction 
of $\sim$3 mag, which is supported by the H$\alpha$
surface brightness variations in the bright 280 km~s$^{-1}$
component seen in the long-slit echelle observations.
Thus we suggest that candidate 2 is associated with a
dust/molecular cloud, which obscures some optical emission
but not the radio emission.
It is not clear whether star formation is taking place locally.
Detailed analyses of the {\it SST} IRAC observations and 
high-resolution near-IR observations are needed to determine
the star formation activity in this molecular cloud.

\acknowledgments

We thank the anonymous referee for critically reading the
manuscript and making helpful comments to improve this 
paper.
We also thank Bernhard Brandl for useful discussion and showing 
us a properly oriented {\it SST} IRAC image of 30 Dor for 
comparison with optical and radio observations.

\begin{figure}
\figurenum{1}
\plotone{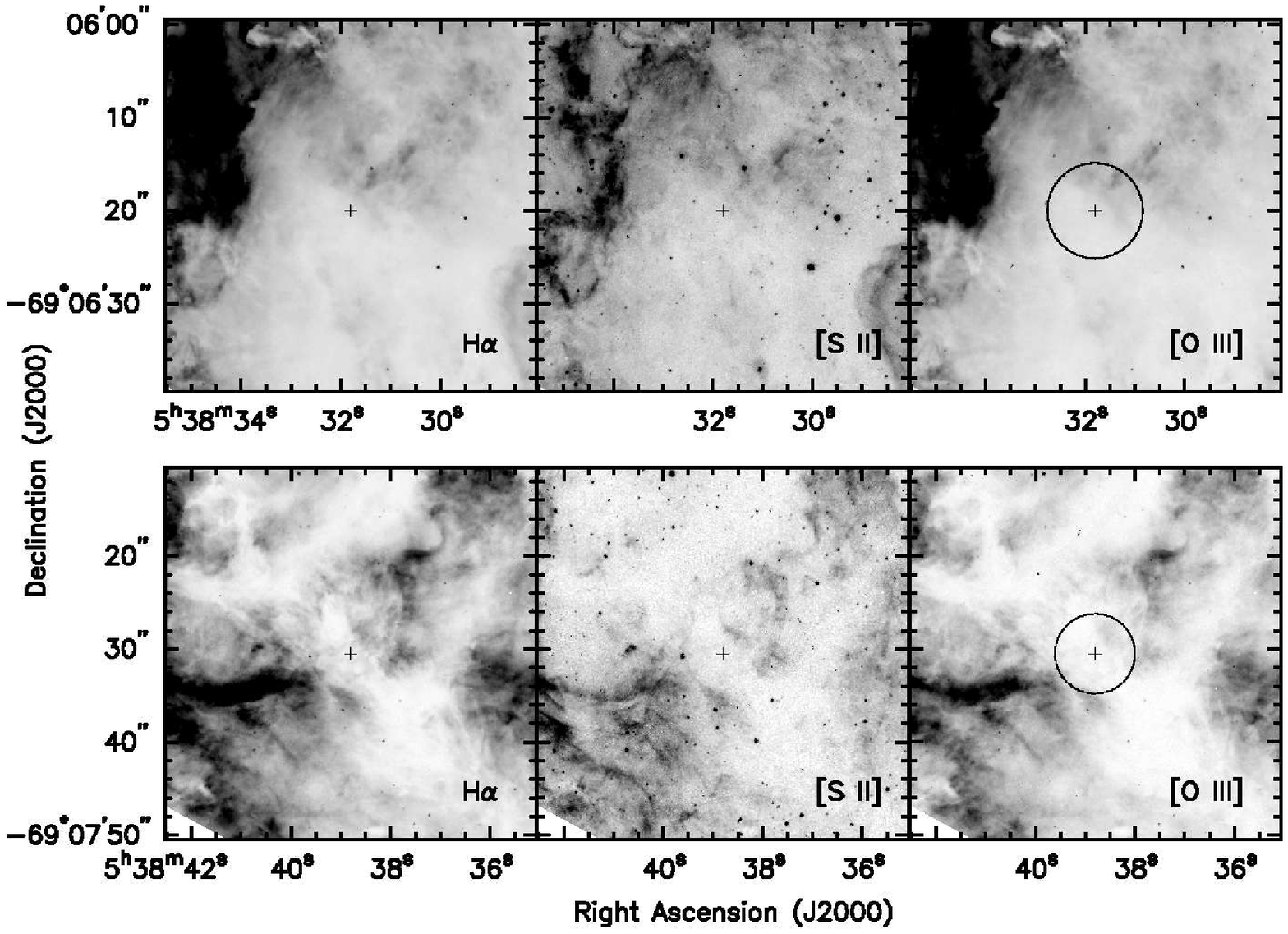}
\caption{{\it HST} WFPC2 images of radio SNR candidates 
MCRX J053831.8$-$690620 (top row; candidate 1) and 
MCRX J053838.8$-$690730 (bottom row; candidate 2) in
the H$\alpha$, [\ion{S}{2}], and [\ion{O}{3}] lines.
The center of the SNR candidate is marked by a cross and
the size marked by a circle.}
\end{figure}

\begin{figure}
\figurenum{2}
\epsscale{0.8}
\plotone{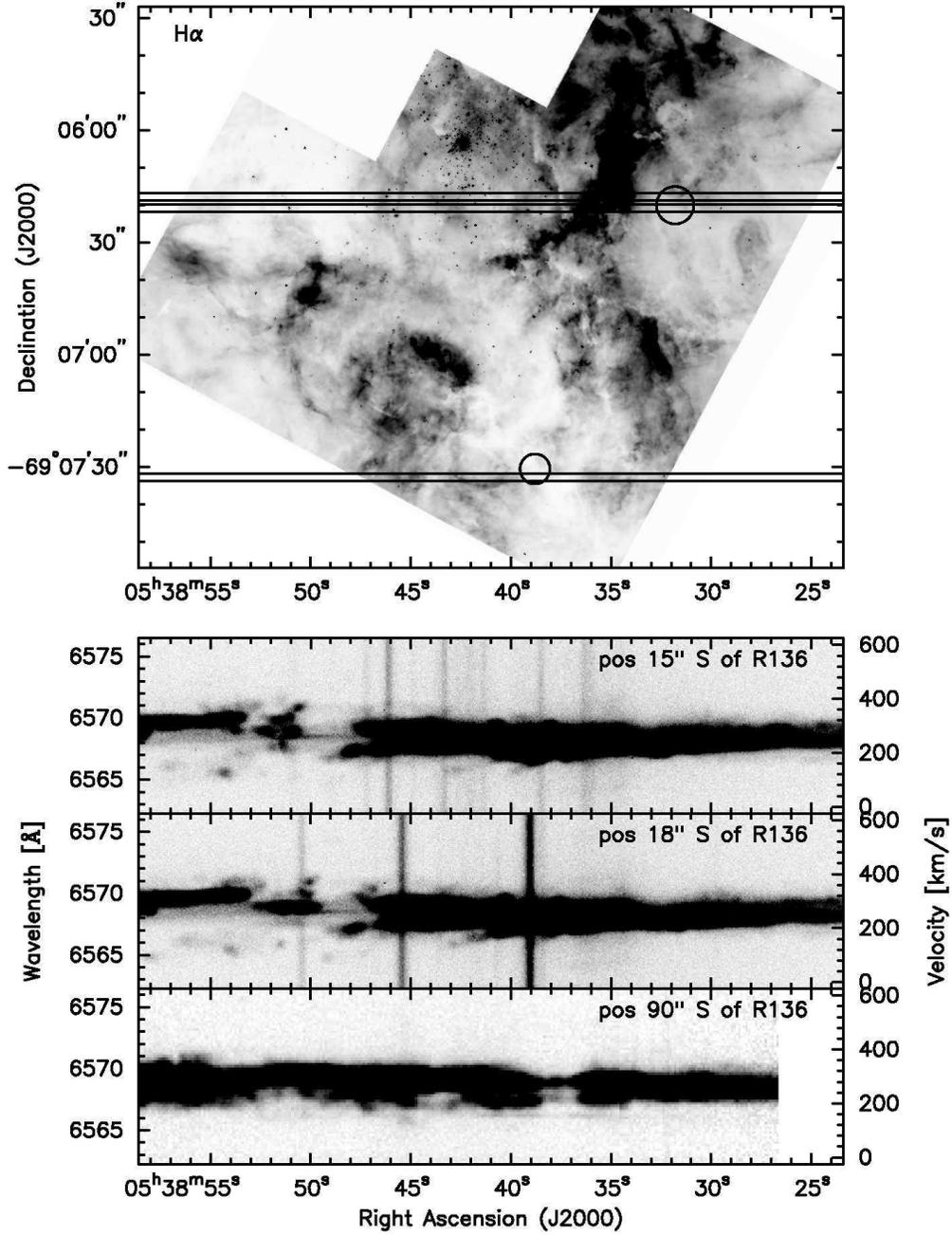}
\caption{Top: {\it HST} WFPC2 H$\alpha$ image of the core of 30 Dor.
The R136 cluster is located near the top center.
The two circles mark the radio SNR candidates suggested by
\citet{LDJ03}.  Bottom: Three echelle spectrograms of the H$\alpha$
line.  The slit positions and widths are marked in the H$\alpha$ 
image above.  The velocities are heliocentric.}
\end{figure}

\begin{deluxetable}{lccc}
\tablewidth{0pc}
\tablecaption{X-ray Luminosity in  the 0.5--10 keV Band}
\tablehead{
Model\,\tablenotemark{a} & $A_V$  &  $L_{\rm X}$ of Cand.\ 1
& $L_{\rm X}$ of Cand.\ 2 \\
      & (mag)  &   (ergs~s$^{-1}$)         &  (ergs~s$^{-1}$)
}
\startdata
RS $kT$=0.5 keV  & 1  & $9.4\times10^{33}$  &  $1.5\times10^{34}$ \\
RS $kT$=1.0 keV  & 1  & $3.9\times10^{33}$  &  $6.0\times10^{33}$ \\
RS $kT$=5.0 keV  & 1  & $3.6\times10^{33}$  &  $6.0\times10^{33}$ \\
RS $kT$=0.5 keV  & 3  & $7.5\times10^{34}$  &  $1.2\times10^{35}$ \\
RS $kT$=1.0 keV  & 3  & $1.5\times10^{34}$  &  $2.4\times10^{35}$ \\
RS $kT$=5.0 keV  & 3  & $6.9\times10^{33}$  &  $1.1\times10^{34}$ \\
RS $kT$=0.5 keV  & 6  & $3.3\times10^{35}$  &  $5.4\times10^{35}$ \\
RS $kT$=1.0 keV  & 6  & $4.2\times10^{34}$  &  $6.9\times10^{34}$ \\
RS $kT$=5.0 keV  & 6  & $1.1\times10^{34}$  &  $1.8\times10^{34}$ \\
RS $kT$=0.5 keV  & 10 & $1.1\times10^{36}$  &  $1.8\times10^{36}$ \\
RS $kT$=1.0 keV  & 10 & $9.6\times10^{34}$  &  $1.5\times10^{35}$ \\
RS $kT$=5.0 keV  & 10 & $1.6\times10^{34}$  &  $2.5\times10^{34}$ \\
PL $\Gamma$=2.0  & 1  & $3.9\times10^{33}$  & $6.3\times10^{33}$ \\
PL $\Gamma$=2.0  & 3  & $7.8\times10^{33}$  & $1.2\times10^{34}$ \\
PL $\Gamma$=2.0  & 6  & $1.3\times10^{34}$  & $2.0\times10^{34}$ \\
PL $\Gamma$=2.0  & 10  & $1.8\times10^{34}$  & $2.9\times10^{34}$ \\
\enddata
\tablenotetext{a}{RS -- \citet{RS77}; PL -- power-law.}
\end{deluxetable}

\end{document}